\documentclass[preprint,3p,11pt,authoryear]{elsarticle}

\usepackage{amssymb}
\usepackage{amsmath}
\usepackage{amsthm}
\allowdisplaybreaks
\newtheorem{proposition}{Proposition}
\usepackage{algorithm}
\usepackage{algorithmicx}
\usepackage{algpseudocode}
\usepackage{setspace}
\usepackage{booktabs}
\usepackage{float}
\usepackage{url}
\usepackage[colorlinks=true]{hyperref}

\usepackage{setspace}
\journal{European Journal of Operational Research}

\begin{document}

\begin{frontmatter}



\title{Movement Prediction-Adjusted Naive Forecast: Is the Naive Baseline Unbeatable in Financial Time Series Forecasting?} 

%

\author{Cheng Zhang\textsuperscript{*}} 

\affiliation{organization={School of Electrical and Electronics Engineering, Hubei Polytechnic University},
            addressline={}, 
            city={Huangshi},
            postcode={435003}, 
            state={Hubei},
            country={China}}
\cortext[cor]{Corresponding author. Email: zhangcheng01@hbpu.edu.cn, ORCID: https://orcid.org/0000-0002-4150-3371, Tel: +86 18271631825}
\begin{abstract}
In financial time series forecasting, the naive forecast is a notoriously difficult benchmark to surpass because of the stochastic nature of the data. Motivated by this challenge, this study introduces the movement prediction-adjusted naive forecast (MPANF), a forecast combination method that systematically refines the naive forecast by incorporating directional information. In particular, MPANF adjusts the naive forecast with an increment formed by three components: the in-sample mean absolute increment as the base magnitude, the movement prediction as the sign, and a coefficient derived from the in-sample movement prediction accuracy as the scaling factor. The experimental results on eight financial time series, using the RMSE, MAE, MAPE, and sMAPE, show that with a movement prediction accuracy of approximately 0.55, MPANF generally outperforms common benchmarks, including the naive forecast, naive forecast with drift, IMA(1,1), and linear regression. These findings indicate that MPANF has the potential to outperform the naive baseline when reliable movement predictions are available.
\end{abstract}



\begin{keyword}
{Finance \sep Forecasting \sep Movement prediction \sep Naive baseline \sep Forecast combination}


\end{keyword}

\end{frontmatter}



\section{Introduction}
Financial time series forecasting plays a crucial role in many decision-making processes. The persistent demand for accurate point forecasts in financial markets has drawn sustained attention from researchers and practitioners. However, although financial time series have long served as testbeds for forecasting methods, their stochastic nature continues to pose a challenge in achieving accurate predictions \citep{Hamilton2020}. One prominent theory on this challenge is the efficient market hypothesis (EMH), which suggests that asset prices incorporate all available information, rendering historical prices unpredictable \citep{Fama1970}. The naive forecast, which predicts the next value of the target variable to be the same as its current value, is frequently regarded as the optimal benchmark for these tasks. 

In support of the EMH, \citet{Moosa2013} reported that as the volatility of financial time series increases, the root mean square error (RMSE) of the forecasting models increases more quickly than that of the naive forecast. Moreover, \citet{Moosa2014} argued that claims of outperforming the naive forecast in terms of the RMSE are often misleading, frequently involving introduced dynamics without rigorous statistical validation. Despite numerous attempts to exploit weak structural signals in financial time series, existing forecasting methods, including statistical, machine learning, and deep learning approaches, have typically failed to outperform the naive baseline \citep{Kilian2003, Moosa2016, Thakkar2021, FotiosPetropoulos2022, Ellwanger2023, Hewamalage2023, Zeng2023, Beck2025}. 

From the forecaster’s perspective, the apparent unpredictability of a variable may stem from an incomplete understanding of the underlying processes, limited knowledge of contributing factors, or limitations in existing analytical methods \citep{Laplace2012}. In this sense, randomness in data may not be entirely inherent but may be partly a reflection of current modeling constraints \citep{Crutchfield2003}. While useful information can be present in the input space, it is often masked by irrelevant signals, which makes it difficult for regression methods to fully capture the informative content \citep{John1994,Guyon2003,Verleysen2005}. Furthermore, identifying the predictors that are most relevant to the target variable remains a complex task \citep{Granger1969,Stock2002}. Nevertheless, these perspectives and empirical findings, while highlighting the practical difficulties of uncovering exploitable patterns, do not preclude the predictability that may exist in the data.

From an information-theoretic perspective, forecasting is theoretically possible when the mutual information between the predictors and the target variable is greater than zero, as this confirms that a nonzero dependency exists \citep{Kraskov2004,Cover2012}. However, practical forecasting feasibility depends on the magnitude, nature, and stability of this mutual information. In binary forecasting tasks, such as predicting whether a value will rise or fall, even modest dependencies can reduce uncertainty and enable models to outperform random guessing \citep{Ircio2020}. In contrast, for continuous point forecasting tasks, a similar magnitude of dependency may only marginally reduce the overall uncertainty of the numerical output \citep{Beraha2019}. This asymmetry partly accounts for the abundance of research on movement prediction in financial time series.

Indeed, previous studies have noted that movement prediction is generally more tractable than point forecasting in financial time series \citep{Taylor2008,Moosa2013}. Behavioral finance studies suggest that investor sentiment and psychological factors can generate predictable movement patterns \citep{De1985,Lee1991,Barberis1998,Baker2007}. Consequently, predictors such as sentiment indicators are often used to improve movement prediction accuracy. For example, \citet{Weng2017} incorporated Google news counts and Wikipedia page views alongside historical data to predict the next day’s movement of AAPL stock via support vector machines (SVMs), achieving up to 0.80 movement prediction accuracy. \citet{Ma2023} proposed a multisource aggregated classification (MAC) method for the prediction of stock price movement, incorporating the numerical features and market-driven news sentiments of target stocks, and the movement prediction accuracy reached 0.70. \citet{Wang2025} used multimodal data that included news data to train a hybrid model consisting of a long short-term memory (LSTM) network and a transformer for stock movement prediction, and the accuracy fell within the range of 0.57--0.87. Depending on the complexity of the markets and availability of data, the movement prediction accuracy of financial time series typically ranges between 0.55 and 0.80 \citep{Bustos2020}.

\begin{figure}[!t]
  \centering
  \includegraphics[width=0.3\textwidth]{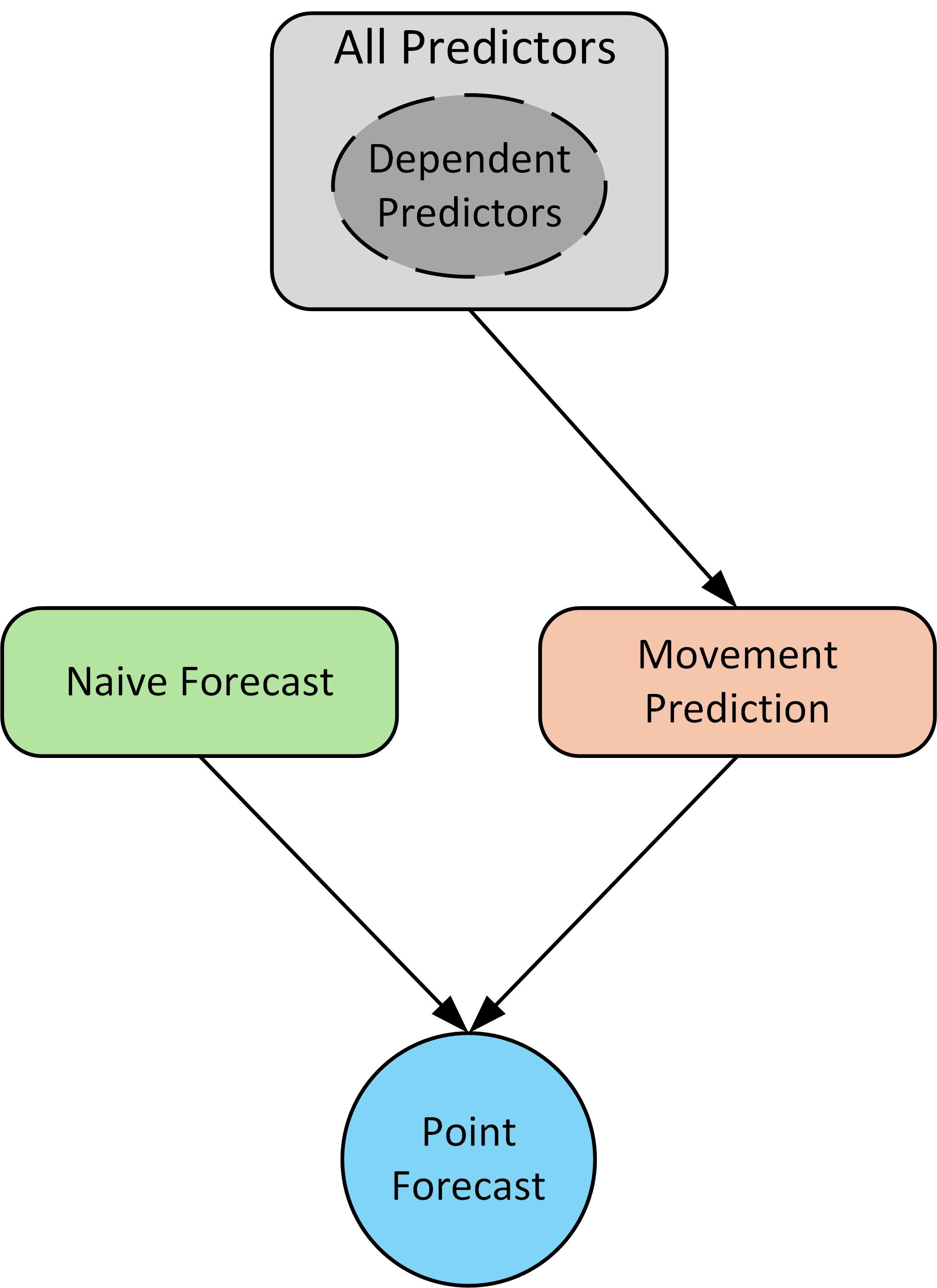}
  \caption{Forecasting framework incorporating naive forecast and movement prediction.}
  \label{fig:fig1}
\end{figure}

Since various studies have demonstrated the feasibility of meaningful movement prediction, which can provide directional information that the naive forecast cannot offer, it is natural to wonder whether such directional information can be exploited to improve the point forecast performance. Guided by the “less is more” principle, which suggests that reducing informational complexity can improve decision-making \citep{Gigerenzer2009}, a method that aims to incorporate directional information for point forecasting can rely solely on naive forecasts and movement predictions as inputs and can be considered a combination of the two types of forecasts. In doing so, the method avoids the potential complications introduced by additional predictors that may harm the final point forecast performance and can serve as the final step in a forecasting framework, as shown in Figure~\ref{fig:fig1}. Unfortunately, existing research has largely treated movement prediction and point forecasting as separate tasks, and most forecast combination methods focus on aggregating forecasts of the same type \citep{Wang2023}. To the best of our knowledge, few studies have examined how and why movement prediction can be integrated with naive forecasts to potentially improve point forecasting performance.

To investigate the potential of combining the movement prediction with the naive forecast for point forecasting, this study proposes a forecast combination method, named the movement prediction-adjusted naive forecast (MPANF). In particular, MPANF is a model-agnostic combination method that can use the output of any classification model and adjust the naive forecast with an increment formed by three components: the in-sample mean absolute increment as the base magnitude, the movement prediction as the sign, and a coefficient derived from the in-sample movement prediction accuracy as the scaling factor. The effectiveness of MPANF is evaluated empirically on eight real-world financial time series from the U.S. market. In all the evaluations, the naive forecast, naive forecast with drift, integrated moving average of order (1,1) (IMA(1,1)), and linear regression serve as baselines, with the RMSE, mean absolute error (MAE), mean absolute percentage error (MAPE) and symmetric mean absolute percentage error (sMAPE) used as evaluation metrics.

This study makes four significant contributions to the literature:
\begin{itemize}
\item First, it introduces a novel forecast combination method that systematically incorporates movement prediction into point forecasting, thereby linking two tasks that have typically been studied separately. While binary classification results have previously been used to guide regression tasks, using the target variable’s own movement prediction as the sole complementary information to enhance the naive forecast provides a fresh methodological perspective. 

\item Second, it provides a theoretical foundation for why combining movement prediction with a naive forecast can generate a point forecast that is superior to the naive baseline. The analytical estimate of the optimal coefficient in the proposed method establishes a connection between movement prediction accuracy and improvement upon the naive baseline, underscoring the informational value of directional signals and distinguishing the proposed method from previous forecasting methods.

\item Third, it derives a retrospective condition, the joint requirement on the out-of-sample coefficient and mean absolute increment, for MPANF to outperform the naive baseline on the out-of-sample set. This condition provides both a theoretical basis and a diagnostic tool, moving beyond the common practice of explaining forecasting performance merely as “overfitting” or “good generalization” by offering a concrete, testable criterion.

\item Fourth, it provides empirical evidence on the conditional predictability of financial time series. For time series that are often regarded as challenging to predict, MPANF can outperform the naive baseline when accurate movement predictions are available.
\end{itemize}

The remainder of this paper is organized as follows. Section 2 presents the details of MPANF, including its theoretical framework and algorithmic implementation. Section 3 reports experimental results on its performance using real-world financial time series. Section 4 discusses the implications of the findings and the limitations of this study. Finally, Section 5 concludes this study and outlines directions for future research.

\section{MPANF}
In this section, we first outline the MPANF's general formulation, followed by the analytical estimate of its optimal coefficient and the retrospective condition that governs its empirical performance. Next, we detail the algorithmic implementation of the MPANF for out-of-sample forecasting.

\subsection{General Formulation}
The value of a time series at time step $t$ can be expressed as the sum of its previous value at time step $t-1$ and an increment ${{\epsilon }_{t}}$ between two adjacent observations. This relationship can be expressed as:

\begin{equation}
\label{eq:1}
{{y}_{t}}={{y}_{t-1}}+{{\epsilon }_{t}},
\end{equation}

where ${y}_{t}$ is the point value at time step $t$ and ${y}_{t-1}$ is the point value at time step $t-1$. The increment $\epsilon_t$ at time step $t$ can be expressed as a product of two terms: a nonnegative term $|\epsilon_{t}|$ representing the magnitude of the increment and a sign term ${d}_{t}$, which indicates the direction of the increment between two adjacent points and takes a value of $+1$ for upward movement and $-1$ for downward movement. This allows ${y}_{t}$ to be reformulated as follows:

\begin{equation}
\label{eq:2}
{{y}_{t}}={{y}_{t-1}}+{{d}_{t}}\cdot |\epsilon_{t}|;|\epsilon_{t}|\geq0
\end{equation}

The nonnegative term $|\epsilon_{t}|$ can be further decomposed into two components as follows: 

\begin{equation}
\label{eq:3}
{{y}_{t}}={{y}_{t-1}}+{{d}_{t}}\cdot {\alpha_t}\cdot \bar{\epsilon}; {\alpha_t}\geq0,\bar{\epsilon }>0,
\end{equation}

where $\bar{\epsilon}$ represents the mean absolute increment of the time series, and $\alpha_t$ represents a nonnegative coefficient that controls the magnitude of the increment at time step $t$. Eq. \eqref{eq:3} implies that the point forecast of the target variable can be generated in such a way that ${{d}_{t}}$ can be predicted and $\alpha_t$ can be estimated. Consider a simplified forecasting strategy in which the increments for each future time step are set to have a fixed magnitude; in other words, we set $\alpha_t$ to be a constant value and derive the one-step-ahead point forecast at time step $t$, denoted by $\hat{y}_t$, as follows:

\begin{equation}
\label{eq:4}
{{\hat{y}}_{t}}={{y}_{t-1}}+\hat{d}_{t}\cdot \alpha\cdot \bar{\epsilon };\bar{\epsilon }>0,
\end{equation}

where ${{\hat{d}}_{t}}$ is the movement prediction, $\alpha$ is a fixed scalar, and $\alpha \in \mathbb{R}$. According to Eq. \eqref{eq:4}, $\hat{y}_t$ can be regarded as an adjustment of the naive forecast with an increment formed by three components: $\bar{\epsilon}$ as the base magnitude, ${{\hat{d}}_{t}}$ as the sign, and $\alpha$ as the scaling factor. As movement prediction plays a central role in enabling this adjustment, the resulting forecast is referred to as the movement-prediction-adjusted naive forecast (MPANF).

\subsection{Optimal Coefficient}
Suppose that the movement prediction of the target variable is provided by an external model. In this case, only the coefficient $\alpha$ in the MPANF requires optimization. When the mean squared error (MSE) is adopted as the cost function, the original optimization problem can be reformulated as an equivalent problem, which can be further approximated by a convex surrogate. Solving this convex approximation yields a closed-form expression for $\alpha$, which we interpret as an analytical estimate of the optimal in-sample coefficient for the MPANF. 

Since the in-sample MSE of the naive forecast is a constant value, minimizing the in-sample MSE of the MPANF is equivalent to maximizing the in-sample MSE difference between the naive forecast and the MPANF:
\begin{equation}
\label{eq:5}
\ minimize\ MSE_{\text{in}}^{\text{MPANF}} \equiv \ maximize \left( MSE_{\text{in}}^{\text{Naive}} - MSE_{\text{in}}^{\text{MPANF}} \right),
\end{equation}

where $MSE_{\text{in}}^{\text{Naive}}$ and $MSE_{\text{in}}^{\text{MPANF}}$ denote the in-sample MSEs of the naive forecast and the MPANF, respectively. Suppose that the in-sample set has $M$ time steps and that the in-sample movement prediction yields $m_{\text{in}}^{\text{correct}}$ correct and $m_{\text{in}}^{\text{incorrect}}$ incorrect directions; then, the in-sample movement prediction accuracy, denoted by $ACC_{\text{in}}$, can be expressed as follows:
\begin{equation}
\label{eq:6}
ACC_{\text{in}}=\frac{m_{\text{in}}^{\text{correct}}}{m_{\text{in}}^{\text{incorrect}}+m_{\text{in}}^{\text{correct}}}
\end{equation}

Assuming that the in-sample movement prediction is unbiased and contains sufficient numbers of both correct and incorrect predictions, the in-sample MSE difference between the naive forecast and the MPANF, denoted by $\Delta MSE_{\text{in}}$, can be approximated as follows:

\begin{equation}
\label{eq:7}
\begin{aligned}
  \Delta MSE_{\text{in}} &= MSE_{\text{in}}^{\text{Naive}} - MSE_{\text{in}}^{\text{MPANF}} \\
  &\approx (4\cdot \alpha_{\text{in}} \cdot ACC_{\text{in}} - \alpha_{\text{in}}^2 - 2\cdot \alpha_{\text{in}}) \cdot \bar{\epsilon}_{\text{in}}^2,    
\end{aligned}
\end{equation}
where $\alpha_{\text{in}}$ denotes the in-sample coefficient $\alpha$ and $\bar{\epsilon}_{\text{in}}$ denotes the in-sample mean absolute increment. A detailed algebraic derivation of the $\Delta MSE_{\text{in}}$ is provided in Appendix A. 

According to Eq. \eqref{eq:7}, maximizing the approximation of $\Delta MSE_{\text{in}}$ is equivalent to maximizing the following concave function of $\alpha_{\text{in}}$:
\begin{equation}
\label{eq:8}
f(\alpha_{\text{in}}) = 4 \cdot \alpha_{\text{in}} \cdot ACC_{\text{in}} - \alpha_{\text{in}}^2 - 2 \cdot \alpha_{\text{in}}
\end{equation}

By differentiating this function with respect to $\alpha_{\text{in}}$ and setting the derivative to zero, we obtain the analytical estimate of the optimal in-sample coefficient, which is formalized in Proposition 1.

\begin{proposition}[\textbf{Coefficient Estimate}]\label{prop:short1}
The analytical estimate of the optimal in-sample coefficient, denoted by $\hat\alpha_{\text{in}}^{*}$, which is the optimal solution for a convex approximation of the in-sample MSE difference between the naive forecast and the MPANF, is given by:
\begin{equation}
\label{eq:9}
\hat\alpha_{\text{in}}^{*} = 2 \cdot ACC_{\text{in}} - 1
\end{equation}
\end{proposition}

Eq. \eqref{eq:9} indicates that this estimate can be derived using only the in-sample movement prediction accuracy and is independent of the characteristics of the in-sample portion of the target series. Substituting this analytical estimate from Eq. \eqref{eq:9} back into the approximated MSE difference expression based on Eq. \eqref{eq:7} yields the theoretical in-sample MSE improvement, as formalized in Proposition 2.

\begin{proposition}[\textbf{In-sample MSE Improvement}]\label{prop:short2}
Using the analytical estimate of the optimal in-sample $\alpha$, the MPANF theoretically achieves a nonnegative MSE improvement over the naive forecast on the in-sample set:
\begin{equation}
\label{eq:10}
\max \Delta MSE_{\text{in}} \approx \left(\hat\alpha_{\text{in}}^{*}\right)^2 \cdot \bar{\epsilon}_{\text{in}}^2 \geq 0
\end{equation}
\end{proposition}

As Proposition 2 demonstrates, when a movement prediction is better than random guessing, a measurable in-sample MSE improvement by the MPANF over the naive baseline is achievable. This improvement is proportional to the movement prediction accuracy: as the accuracy increases, the in-sample MSE improvement also increases. At a movement prediction accuracy of 1.0, the maximum achievable in-sample MSE improvement is approximately $\bar{\epsilon}_{\text{in}}^2$. Thus, under idealized conditions, the MPANF's expected performance in terms of the MSE matches or surpasses that of the naive forecast on the in-sample set:
\begin{equation}
\label{eq:11}
MSE_{\text{in}}^{\text{MPANF}} \leq MSE_{\text{in}}^{\text{Naive}}
\end{equation}

\subsection{Retrospective Condition}

Suppose that $\hat\alpha_{\text{in}}^{*}$ and $\bar{\epsilon}{\text{in}}$ are adopted by the MPANF for the out-of-sample forecast, and that the out-of-sample set contains $N$ time steps, with the out-of-sample movement prediction yielding $n_{\text{out}}^{\text{correct}}$ correct and $n_{\text{out}}^{\text{incorrect}}$ incorrect predictions; then, the out-of-sample movement prediction accuracy, denoted by $ACC_{\text{out}}$, can be expressed as:

\begin{equation}
\label{eq:12}
ACC_{\text{out}}=\frac{n_{\text{out}}^{\text{correct}}}{n_{\text{out}}^{\text{incorrect}}+n_{\text{out}}^{\text{correct}}}
\end{equation}

The out-of-sample MSE difference between the naive forecast and the MPANF, denoted by $\Delta MSE_{\text{out}}$, can subsequently be approximated as follows:
\begin{equation}
\label{eq:13}
\begin{aligned}
  \Delta MSE_{\text{out}} &= MSE_{\text{out}}^{\text{Naive}} - MSE_{\text{out}}^{\text{MPANF}} \\
   &\approx 2\cdot \hat\alpha_{\text{in}}^{*} \cdot \bar{\epsilon}_{\text{in}} \cdot \hat\alpha_{\text{out}}^{*} \cdot \bar{\epsilon}_{\text{out}} -\left(\hat\alpha_{\text{in}}^{*}\right)^2 \cdot \bar{\epsilon}_{\text{in}}^2,
\end{aligned}
\end{equation}

where $\bar{\epsilon}_{\text{out}}$ denotes the out-of-sample mean absolute increment and $\hat\alpha_{\text{out}}^{*}$ denotes the analytical estimate of the optimal out-of-sample coefficient, which is derived according to Proposition 1:

\begin{equation}
\label{eq:14}
\hat\alpha_{\text{out}}^{*} = 2 \cdot ACC_{\text{out}} - 1
\end{equation}

A detailed algebraic expansion of the $\Delta MSE_{\text{out}}$ is provided in Appendix B.
For the MPANF to outperform the naive forecast in terms of the out-of-sample MSE, the approximated $\Delta MSE_{\text{out}}$ should be no less than zero, which leads to a retrospective condition for out-of-sample outperformance, as formalized in Proposition 3. 

\begin{proposition}[\textbf{Retrospective Condition}]\label{prop:short3}
The retrospective condition for the MPANF to achieve a lower out-of-sample MSE than the naive forecast is given by:

\begin{equation}
\label{eq:15}
\hat\alpha_{\text{out}}^{*} \cdot \bar{\epsilon}_{\text{out}}\geq\frac{1}{2}\cdot\hat\alpha_{\text{in}}^{*} \cdot \bar{\epsilon}_{\text{in}}
\end{equation}
\end{proposition}

Since $\hat\alpha_{\text{out}}^{*}$ and $\bar{\epsilon}_{\text{out}}$ are unknown before the out-of-sample period is observed, whether the condition in Eq. \eqref{eq:15} holds can only be regarded as a probabilistic event. Once the out-of-sample realizations are available, however, the inequality is deterministically evaluated: if Eq. \eqref{eq:15} is satisfied, the MPANF yields a lower out-of-sample MSE than the naive forecast; otherwise, no improvement occurs. According to Eq. \eqref{eq:13}, the magnitude of improvement depends on how closely 
$\hat{\alpha}_{\text{in}}^{*}$ and $\hat{\bar{\epsilon}}_{\text{in}}$ closely match their out-of-sample counterparts. 

\subsection{Algorithmic Implementation}
The MPANF is a model-agnostic forecast combination method that can use movement predictions from any external model. This modularity allows the MPANF to integrate advancements in the field of time series movement prediction without altering its core framework. If the movement prediction is available, the practical implementation of the MPANF in a one-step-ahead forecasting scenario is summarized in Algorithm \ref{alg:1}.
\begin{algorithm}[!t]
\caption{Movement prediction–adjusted naive forecast}
\label{alg:1}
\begin{algorithmic}
\setstretch{1.2}
\Procedure{MPANF}{$Y_{1:M},\ \hat{d}^{\text{in}}_{2:M},\ \{\hat{d}^{\text{out}}_{t+1}\}_{t\ge M}$}
    \State $P \gets [\,]$
    \State $\bar{\epsilon}_{\text{in}} \gets \dfrac{1}{M-1}\sum_{t=2}^{M} \lvert Y_t - Y_{t-1}\rvert$
    \State $d^{\text{in}}_t \gets \mathrm{sign}(Y_t - Y_{t-1})\ \ \text{for } t=2,\dots,M$
    \State $ACC_{\text{in}} \gets \dfrac{1}{M-1}\sum_{t=2}^{M} \mathbf{1}\!\big(\hat{d}^{\text{in}}_t = d^{\text{in}}_t\big)$
    \State $\hat\alpha_{\text{in}}^{*} \gets 2\cdot ACC_{\text{in}} - 1$
    \State $Y_{\mathrm{last}} \gets Y_M$;\ \ $t \gets M$
    \While{\text{next } $\hat{d}^{\text{out}}_{t+1}$ \text{ is available}}
        \State $\Delta \gets \hat{d}^{\text{out}}_{t+1} \cdot\hat\alpha_{\text{in}}^{*} \cdot \bar{\epsilon}_{\text{in}}$
        \State $\hat{Y}_{t+1} \gets Y_{\mathrm{last}} + \Delta$
        \State \textbf{append} $\hat{Y}_{t+1}$ \text{ to } $P$
        \State \textbf{observe} $Y_{t+1}$
        \State $Y_{\mathrm{last}} \gets Y_{t+1}$;\ \ $t \gets t+1$
    \EndWhile
    \State \textbf{return} $P$
\EndProcedure
\Statex
\end{algorithmic}
\end{algorithm}

Algorithm \ref{alg:1} first computes the in-sample mean absolute increment, $\bar{\epsilon}_{\text{in}}$, and the in-sample movement prediction accuracy, $ACC_{\text{in}}$, from the provided in-sample data. These values are then used to calculate the analytically derived coefficient, $\hat\alpha_{\text{in}}^{*}$. In the out-of-sample forecasting stage, the algorithm iteratively generates forecasts $\hat{Y}_{t+1}$ by adjusting the most recent observation $Y_{\text{last}}$. Each adjustment, denoted by $\Delta$, equals the product of $\hat{d}^{\text{out}}_{t+1}$, $\hat\alpha_{\text{in}}^{*}$, and $\bar{\epsilon}_{\text{in}}$. This iterative process produces the set of out-of-sample forecasts.

\section{Experiments}\label{sec-experiments}

This section details the experiments for evaluating the performance of the MPANF using real world data. We first present the experimental setup, including the characteristics of the datasets, the baselines, and the evaluation metrics. Subsequently, we present and analyze the experimental results, which include a comparative assessment of out-of-sample performance, and a validation of the retrospective condition.

\subsection{Experimental Setup}
\subsubsection{Datasets and Preprocessing}

Because the MPANF’s effectiveness depends on the quality of the given movement prediction, data selection should account for both the target series and the corresponding movement prediction availability. Given the well-documented strong co-movement between the Financial Times Stock Exchange (FTSE) index and U.S. stock prices \citep{Sarwar2020,Gupta2024}, together with the fact that the FTSE open price is recorded approximately 13 hours before the U.S. market close price on the same date, we selected eight U.S. financial time series as the target series, including Apple (AAPL), Boeing (BA), Google (GOOGL), Goldman Sachs (GS), Halliburton (HAL), Brent crude oil (OIL), Schlumberger (SLB), and Tesla (TSLA), on the basis of their strong predictive alignment with FTSE movements. Accordingly, the FTSE open price is used as the exogenous variable for generating movement predictions. The details of the movement prediction method, which is based on the co-movement of the target variable and an exogenous variable, are provided in Appendix C.

\begin{figure}[!t]
  \centering
  \includegraphics[width=1\textwidth]{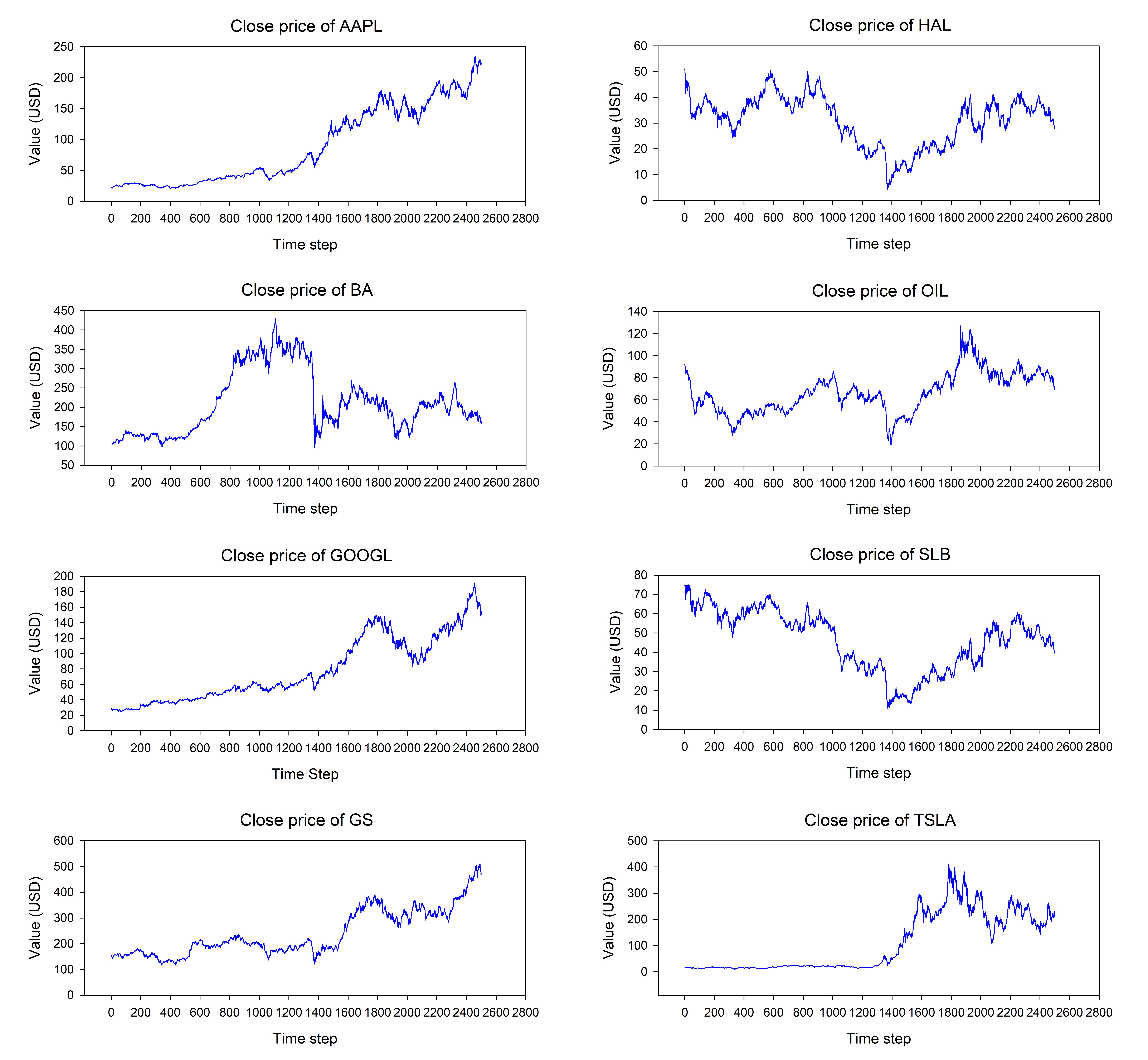}
  \caption{Time series of eight target variables.}
  \label{fig:fig2}
\end{figure}

\begin{figure}[!t]
  \centering
  \includegraphics[width=0.8\textwidth]{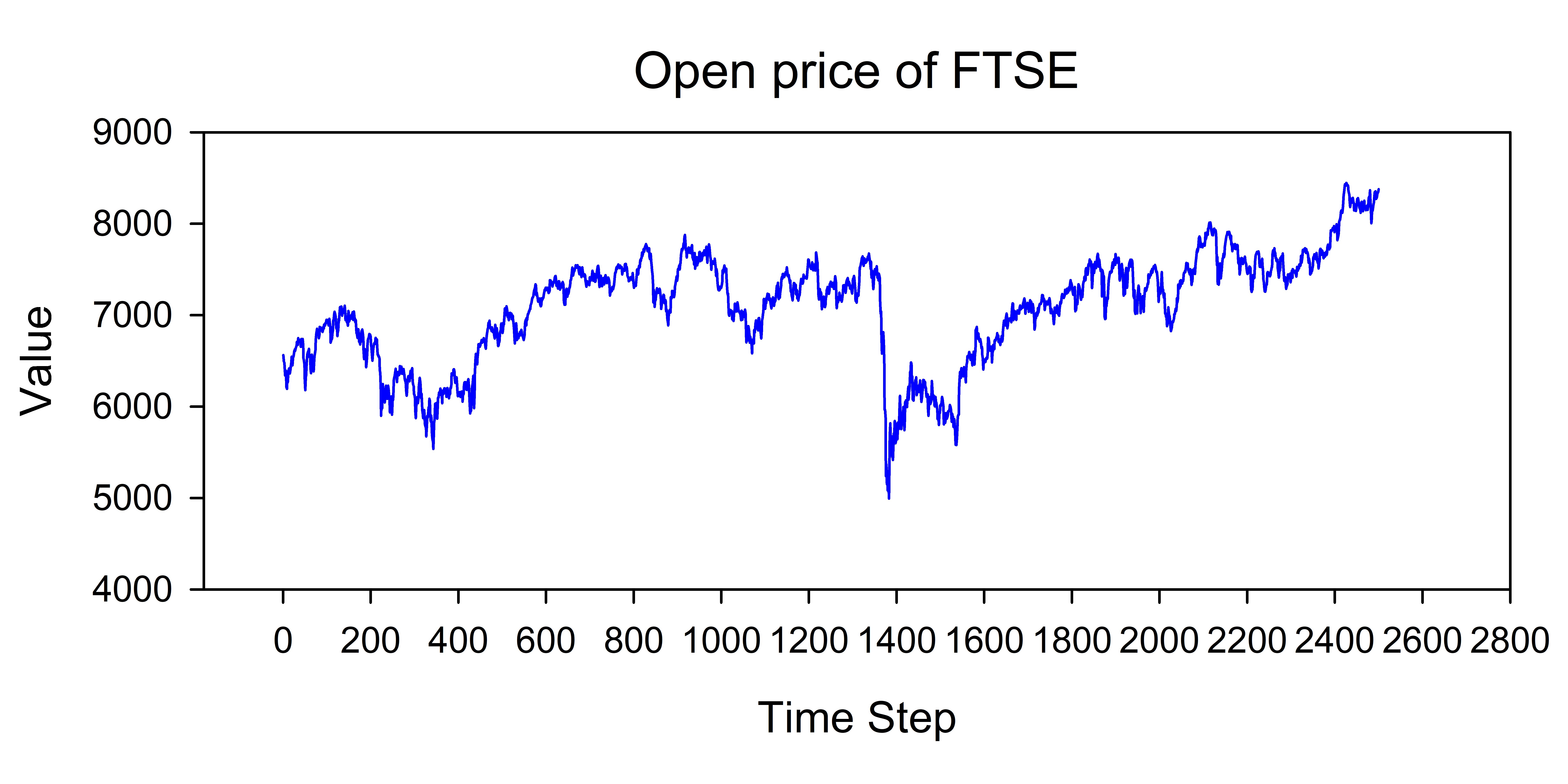}
  \caption{Open price of FTSE.}
  \label{fig:fig3}
\end{figure}

Historical data for the FTSE open price and the close prices of the eight target series were obtained from Yahoo Finance, covering the period from October 8, 2014, to September 13, 2024. The FTSE data were aligned with each target series on a day-by-day basis, with extra observations removed and missing values imputed via forward filling \citep{Junninen2004}. After data alignment, all time series were truncated to a uniform length of 2,500 observations and then split into in-sample and out-of-sample sets at a 1:1 ratio, ensuring sufficient data for both model estimation and evaluation. The close prices of the eight target variables and the FTSE open price are illustrated in Figures \ref{fig:fig2} and \ref{fig:fig3}, respectively, with a brief description provided in Table \ref{table:1}. Using the movement of the FTSE open price as the movement prediction for all eight target series, the in-sample movement prediction accuracy for each dataset exceeded 0.55, indicating that the directional signals derived from the FTSE open price are meaningful. For each target series, the out-of-sample forecasts by the MPANF were produced via Algorithm \ref{alg:1}.

\begin{table}[!t]
\centering
\caption{Statistics of nine financial time series.}
\label{table:1}
\begin{tabular}{@{\hskip 2pt}l@{\hskip 4pt}c@{\hskip 2pt}c@{\hskip 2pt}c@{\hskip 2pt}c@{\hskip 2pt}c@{\hskip 2pt}c@{\hskip 2pt}}
\toprule
\textbf{Time series} & \textbf{Count} & \textbf{Min} & \textbf{Median} & \textbf{Max} & $\mathbf{ACC}_{\text{in}}$ & $\mathbf{\bar{\epsilon}}_{\text{in}}$ \\
\midrule
AAPL (Close price)  & 2500 &  20.72  &  54.16  & 234.55 & 0.5536 & 0.3842 \\
BA (Close price)    & 2500 &  95.01  & 195.92  & 430.30 & 0.5568 & 2.5867 \\
GOOGL (Close price) & 2500 &  24.79  &  61.44  & 190.93 & 0.5504 & 0.4624 \\
GS (Close price)    & 2500 & 117.58  & 202.27  & 510.25 & 0.5504 & 1.9030 \\
HAL (Close price)   & 2500 &   4.32  &  33.30  & 151.09 & 0.5680 & 0.5252 \\
OIL (Close price)   & 2500 &  19.33  &  65.20  & 127.98 & 0.5576 & 0.9067 \\
SLB (Close price)   & 2500 &  11.02  &  49.49  &  74.91 & 0.5536 & 0.6596 \\
TSLA (Close price)  & 2500 &   9.58  &  23.72  & 409.97 & 0.5536 & 0.3534 \\
FTSE (Open price)   & 2500 & 4993.90 & 7210.00 & 8445.80 & --     & --     \\
\bottomrule
\end{tabular}
\end{table}

\subsubsection{Baselines and Evaluation Metrics}

As improvements over naive forecasts in financial time series forecasting can be difficult to achieve, even with sophisticated approaches such as machine learning and deep learning, we adopt simple and widely recognized baselines, including the naive forecast, naive forecast with drift, IMA(1,1), and linear regression, for performance comparison \citet{AndreiRogers1995,Jadevicius2015,Franses2020}. In particular, the linear regression model, which acts as a meta-learner to combine the naive forecast and the movement prediction, uses only the current observed value and the predicted movement as predictors. 

The performance of each method was evaluated via four metrics: the RMSE, MAE, MAPE, and sMAPE. The RMSE, derived from the square root of the MSE, emphasizes larger errors, which makes it suitable for identifying significant deviations in forecasts. The MAE, on the other hand, provides an interpretable measure of the average error without overpenalizing outliers. The MAPE and sMAPE normalize the errors relative to the actual values, enabling comparisons between datasets with different scales. Together, these metrics capture both absolute and relative error characteristics, ensuring a robust and balanced evaluation. The formulas for these metrics are as follows:
\begin{equation}
\label{eq:16}
RMSE=\sqrt{\frac{1}{N}\sum\limits_{i=1}^{N}{{{({{y}_{i}}-{{{\hat{y}}}_{i}})}^{2}}}},
\end{equation}
\begin{equation}
\label{eq:17}
MAE=\frac{1}{N}\sum\limits_{i=1}^{N}{\left| {{y}_{i}}-{{{\hat{y}}}_{i}} \right|},
\end{equation}
\begin{equation}
\label{eq:18}
MAPE=\frac{100}{N}\sum\limits_{i=1}^{N}{\left| \frac{{{y}_{i}}-{{{\hat{y}}}_{i}}}{{{y}_{i}}} \right|},
\end{equation}
\begin{equation}
\label{eq:19}
sMAPE=\frac{100}{N}\sum\limits_{i=1}^{N}{\frac{\left| {{y}_{i}}-{{{\hat{y}}}_{i}} \right|}{\frac{\left| {{y}_{i}} \right|+\left| {{{\hat{y}}}_{i}} \right|}{2}}},
\end{equation}
where $N$ denotes the size of the out-of-sample set, ${{\hat{y}}_{i}}$ denotes the predicted value, and ${{y}_{i}}$ denotes the actual value.

\subsection{Experimental Results}

\subsubsection{Out-of-Sample Performance}

\begin{figure}[!t]
  \centering
  \includegraphics[width=1\textwidth]{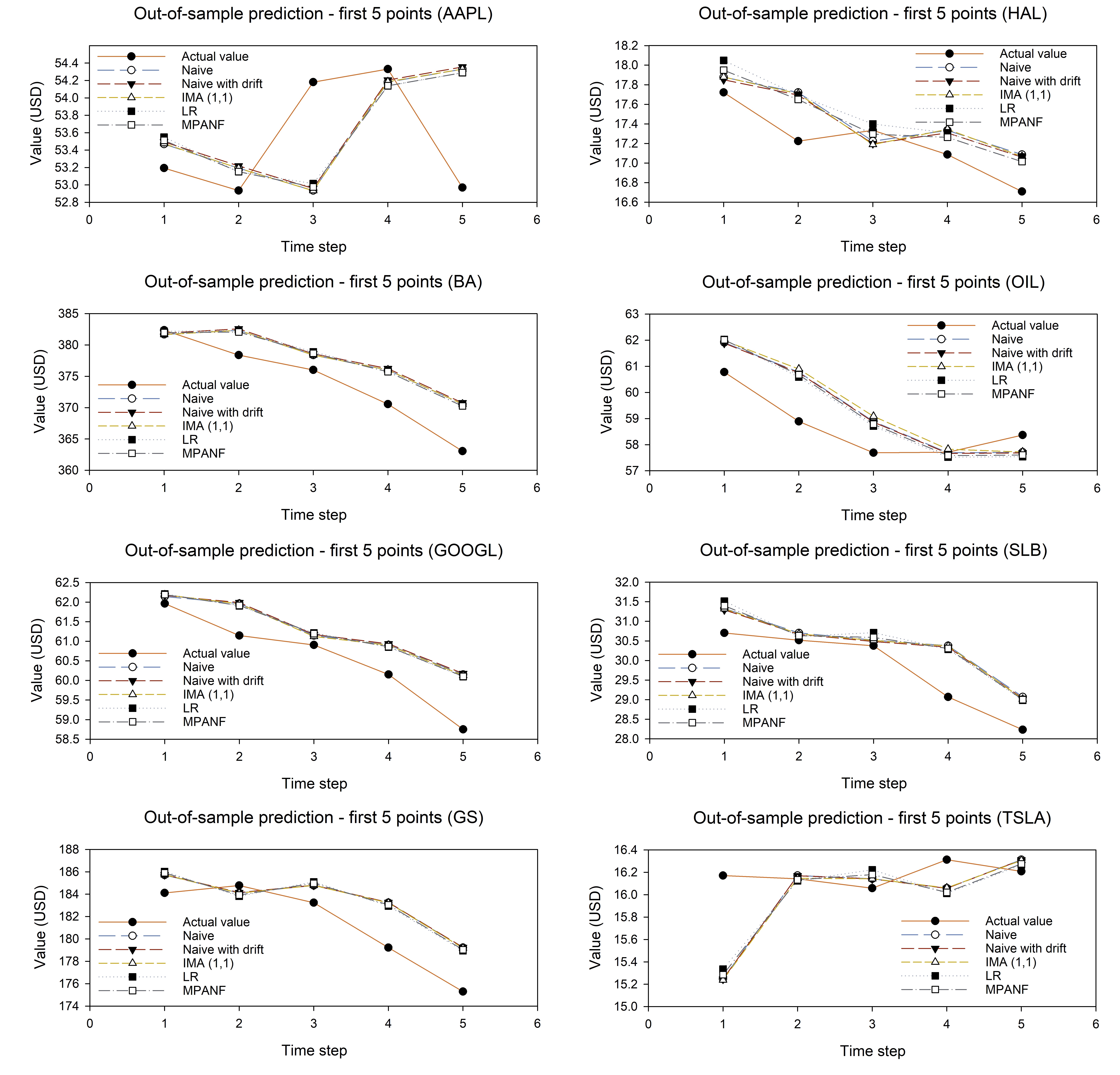}
  \caption{Out-of-sample forecasts by different methods.}
  \label{fig:fig4}
\end{figure}

The out-of-sample forecasting results for the eight financial time series are illustrated in Figure \ref{fig:fig4}. Given that each out-of-sample set consists of 1250 time steps and that predictions from different baseline methods visually overlap to a large extent, only the first five predicted points from each set are displayed for clarity. Small differences between the MPANF and the baseline methods are observed because the overall in-sample movement prediction accuracy for each time series is just slightly above 0.55, leading to a small coefficient $\alpha$, which means that the magnitude of adjustment for each time step is  approximately one tenth of the base magnitude. Moreover, the naive forecast, naive forecast with drift, and IMA(1,1) forecasts yield similar predictions, reinforcing the suitability of these methods as baselines and the robustness of the naive forecast as the benchmark for time series forecasting. Moreover, because the linear regression model closely mirrors the structure of the MPANF, its performance is similarly dependent on movement prediction accuracy and therefore remains close to that of the MPANF.  

Tables \ref{table:2}, \ref{table:3}, \ref{table:4}, and \ref{table:5} present performance comparisons between the MPANF and the baseline methods for eight out-of-sample sets, reported in terms of the RMSE, MAE, MAPE, and sMAPE, respectively. Notably, the MPANF consistently outperforms the naive forecast across all four metrics, except for the MAPE and sMAPE on the GOOGL dataset, where its performance is slightly inferior. Specifically, the MPANF achieves the lowest RMSE in 4 out of 8 cases, the lowest MAE in 6 out of 8 cases, the lowest MAPE in 6 out of 8 cases, and the lowest sMAPE in 5 out of 8 cases, demonstrating its overall superiority over the baseline methods. On average, the MPANF achieved the following improvements over the simple baseline model: a 0.32\%  reduction in the RMSE, a 0.41\%  reduction in the MAE, a 0.47\%  reduction in the MAPE, and a 0.46\%  reduction in the sMAPE.

\begin{table}[!t]
\centering
\caption{RMSE comparison across forecasting methods for selected stocks.}
\begin{tabular}{lccccc}
\hline
\textbf{Stock} & \textbf{Naive} & \textbf{Naive with Drift} & \textbf{IMA(1,1)} & \textbf{LR} & \textbf{MPANF} \\ \hline
AAPL & 2.5937 & \textbf{2.5925} & 2.5979 & 2.5942 & 2.5931 \\
BA   & 5.9080 & 5.9187 & 5.9166 & 5.8932 & \textbf{5.8829} \\
GOOGL& 2.2093 & \textbf{2.2085} & 2.2108 & 2.2114 & 2.2088 \\
GS   & 5.3671 & 5.3660 & 5.3727 & 5.4039 & \textbf{5.3485} \\
HAL  & 0.7334 & 0.7341 & 0.7332 & \textbf{0.7261} & 0.7271 \\
OIL  & 1.8909 & 1.8912 & 1.8976 & 1.8940 & \textbf{1.8864} \\
SLB  & 0.9401 & 0.9410 & 0.9401 & \textbf{0.9331} & 0.9343 \\
TSLA & 8.1524 & 8.1525 & 8.1677 & 8.3351 & \textbf{8.1515} \\ 
\hline
\end{tabular}
\label{table:2}
\end{table}

\begin{table}[!t]
\centering
\caption{MAE comparison across forecasting methods for selected stocks.}
\begin{tabular}{lccccc}
\hline
\textbf{Stock} & \textbf{Naive} & \textbf{Naive with Drift} & \textbf{IMA(1,1)} & \textbf{LR} & \textbf{MPANF} \\ \hline
AAPL & 1.8950 & \textbf{1.8935} & 1.8975 & 1.8958 & 1.8943 \\
BA   & 3.9691 & 3.9821 & 3.9750 & 3.9657 & \textbf{3.9533} \\
GOOGL& 1.5680 & \textbf{1.5660} & 1.5727 & 1.5746 & 1.5674 \\
GS   & 3.9743 & 3.9737 & 3.9811 & 4.0021 & \textbf{3.9534} \\
HAL  & 0.5423 & 0.5429 & 0.5439 & 0.5376 & \textbf{0.5368} \\
OIL  & 1.2959 & 1.2981 & 1.3000 & 1.3073 & \textbf{1.2903} \\
SLB  & 0.6901 & 0.6910 & 0.6907 & 0.6852 & \textbf{0.6843} \\
TSLA & 5.5732 & 5.5733 & 5.5775 & 5.8155 & \textbf{5.5722} \\
\hline
\end{tabular}
\label{table:3}
\end{table}

\begin{table}[!t]
\centering
\caption{MAPE (\%) comparison across forecasting methods for selected stocks.}
\begin{tabular}{lccccc}
\hline
\textbf{Stock} & \textbf{Naive} & \textbf{Naive with Drift} & \textbf{IMA(1,1)} & \textbf{LR} & \textbf{MPANF} \\ \hline
AAPL & 1.4081 & 1.4069 & \textbf{1.4065} & 1.4079 & 1.4074 \\
BA   & 2.0601 & 2.0675 & 2.0633 & 2.0585 & \textbf{2.0514} \\
GOOGL& 1.4357 & \textbf{1.4340} & 1.4380 & 1.4406 & 1.4359 \\
GS   & 1.4037 & 1.4036 & 1.4046 & 1.4074 & \textbf{1.3967} \\
HAL  & 2.3517 & 2.3532 & 2.3667 & 2.3277 & \textbf{2.3216} \\
OIL  & 1.8379 & 1.8409 & 1.8430 & 1.8456 & \textbf{1.8296} \\
SLB  & 2.1522 & 2.1532 & 2.1574 & 2.1355 & \textbf{2.1291} \\
TSLA & 2.9511 & 2.9513 & 2.9549 & 3.0534 & \textbf{2.9509} \\ 
\hline
\end{tabular}
\label{table:4}
\end{table}

\begin{table}[!t]
\centering
\caption{sMAPE (\%) comparison across forecasting methods for selected stocks.}
\begin{tabular}{lccccc}
\hline
\textbf{Stock} & \textbf{Naive} & \textbf{Naive with Drift} & \textbf{IMA(1,1)} & \textbf{LR} & \textbf{MPANF} \\ \hline
AAPL & 1.4088 & \textbf{1.4072} & 1.4072 & 1.4088 & 1.4081 \\
BA   & 2.0542 & 2.0591 & 2.0591 & 2.0500 & \textbf{2.0457} \\
GOOGL& 1.4352 & \textbf{1.4331} & 1.4375 & 1.4410 & 1.4354 \\
GS   & 1.4035 & 1.4032 & 1.4045 & 1.4090 & \textbf{1.3966} \\
HAL  & 2.3377 & 2.3427 & 2.3537 & \textbf{2.3051} & 2.3084 \\
OIL  & 1.8290 & 1.8328 & 1.8336 & 1.8377 & \textbf{1.8210} \\
SLB  & 2.1465 & 2.1503 & 2.1521 & 2.1260 & \textbf{2.1240} \\
TSLA & 2.9518 & 2.9520 & 2.9554 & 3.0767 & \textbf{2.9515} \\
\hline
\end{tabular}
\label{table:5}
\end{table}

In addition, baseline methods such as the naive forecast with drift, IMA(1,1), or linear regression have demonstrated superior performance in certain time series, yielding lower error metric values than other methods do. For example, they produced lower RMSE values for AAPL, GOOGL, HAL, and OIL, and lower MAEs, MAPEs, and sMAPEs for AAPL and GOOGL. However, the performances of these methods are inconsistent; they frequently fall short of the naive forecasts in other time series.

Compared with the naive forecast, linear regression can sometimes perform better, particularly when movement prediction is used as an input feature. However, this occasional improvement lacks a clear theoretical justification, and its performance generally falls short of the MPANF because of a higher susceptibility to overfitting. While linear regression can fit in-sample data closely, its out-of-sample performance degrades significantly if the accuracy of the movement prediction decreases. In contrast, the coefficient in the MPANF is analytically derived rather than obtained through numerical minimization of in-sample errors. As an analytical estimate of the optimal value, the coefficient $\hat{\alpha}_{\text{in}}^{*}$ may deviate slightly from the true empirical optimum. This controlled deviation constrains the model’s flexibility, thereby reducing the risk of overfitting to in-sample noise and enhancing the stability of out-of-sample performance.

\begin{figure}[!t]
  \centering
  \includegraphics[width=1\textwidth]{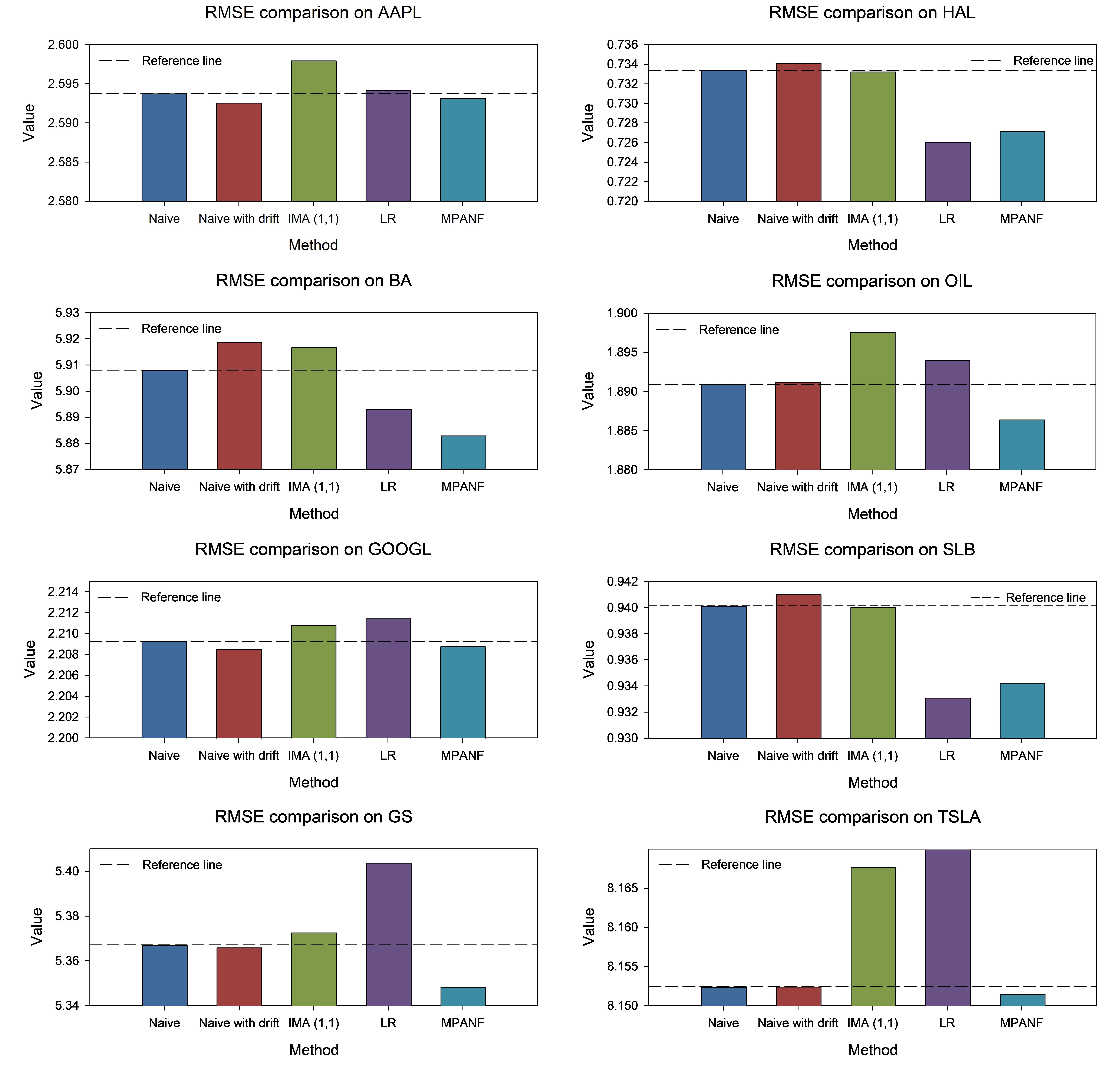}
  \caption{Comparison of the RMSEs of different methods.}
  \label{fig:fig5}
\end{figure}

\begin{figure}[!t]
  \centering
  \includegraphics[width=1\textwidth]{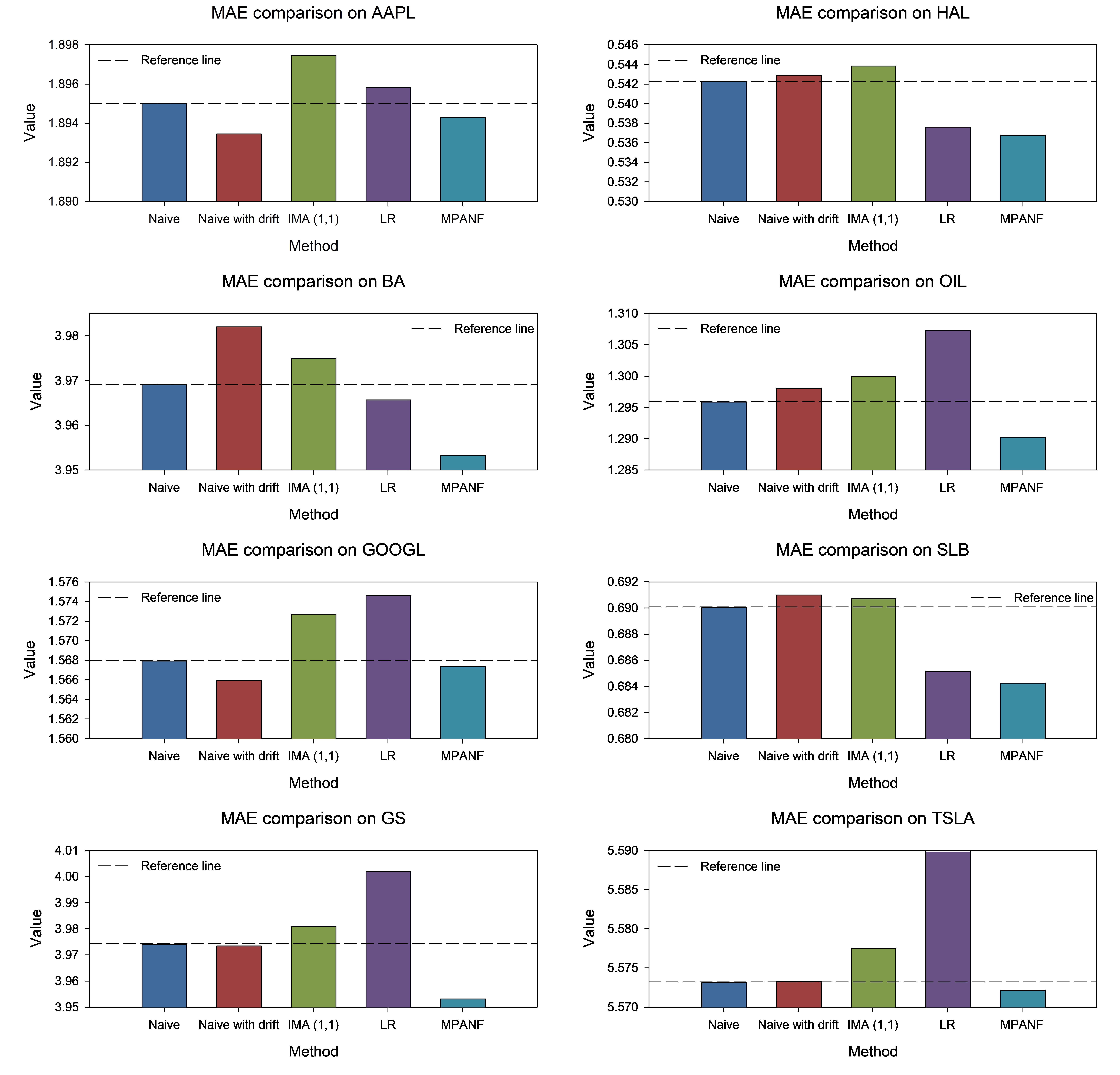}
  \caption{Comparison of the MAEs of different methods.}
  \label{fig:fig6}
\end{figure}

\begin{figure}[!t]
  \centering
  \includegraphics[width=1\textwidth]{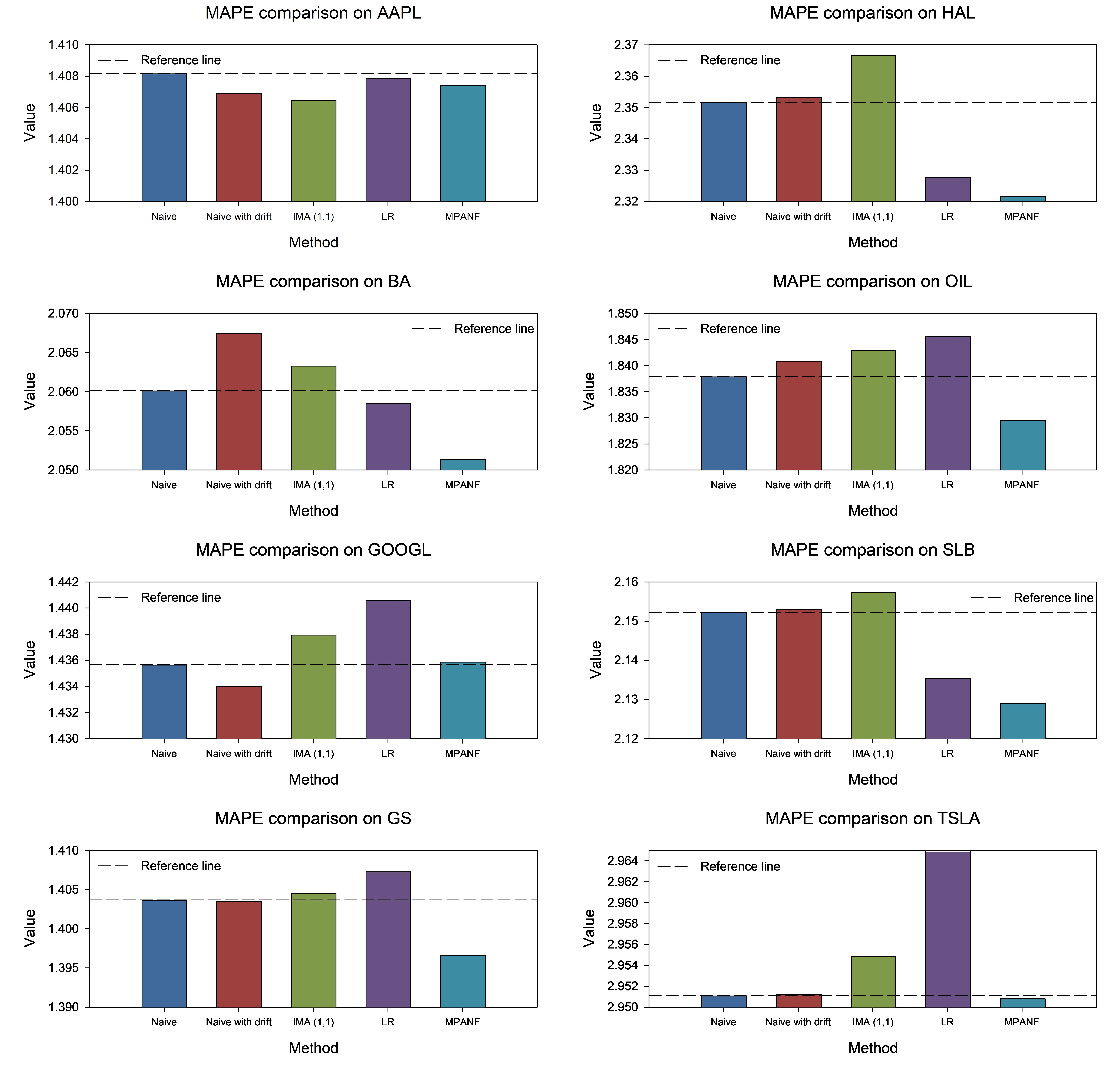}
  \caption{Comparison of the MAPEs of different methods.}
  \label{fig:fig7}
\end{figure}

\begin{figure}[!t]
  \centering
  \includegraphics[width=1\textwidth]{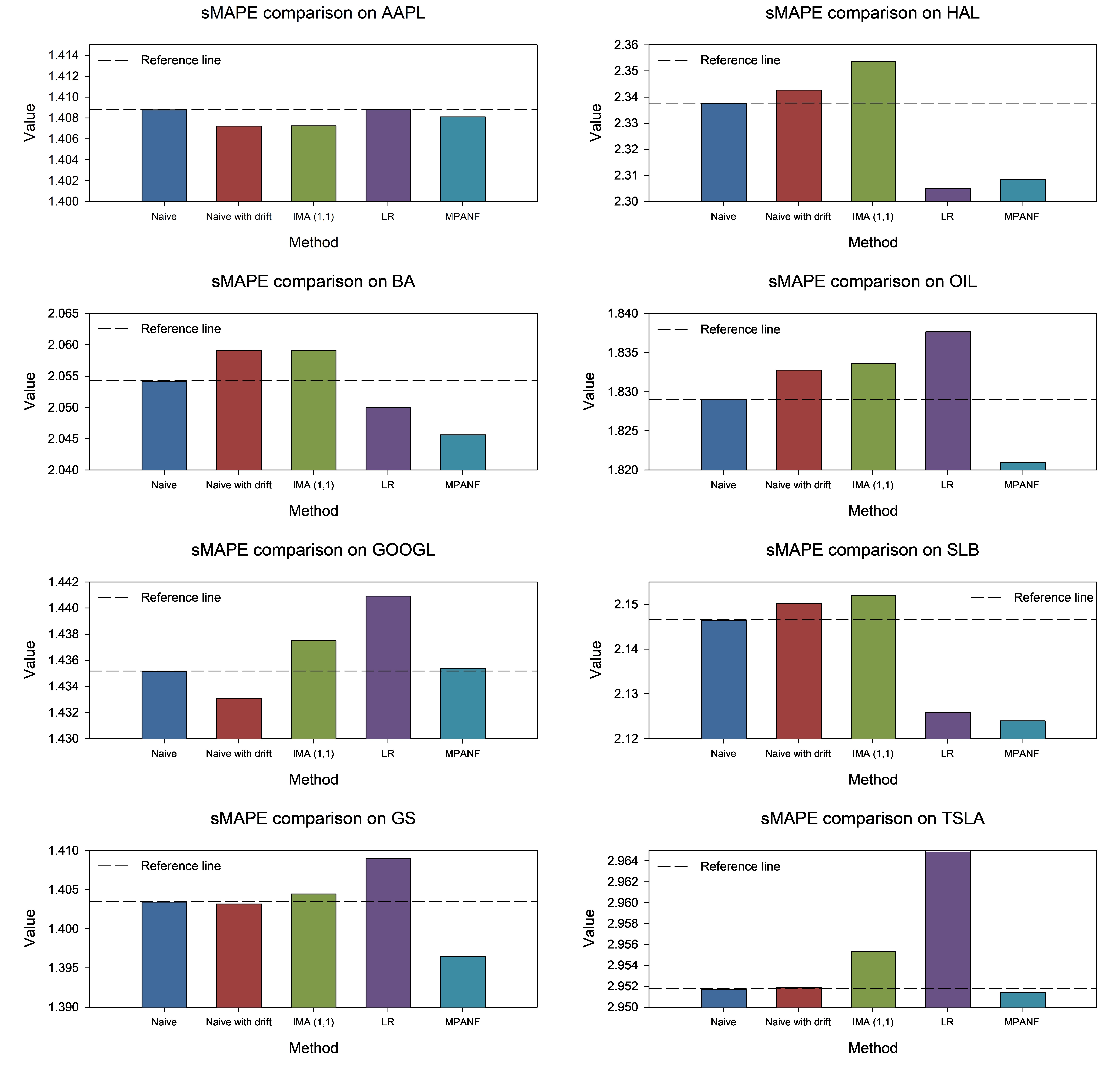}
  \caption{Comparison of the sMAPEs of different methods.}
  \label{fig:fig8}
\end{figure}

As shown in Figures \ref{fig:fig5}, \ref{fig:fig6}, \ref{fig:fig7}, and \ref{fig:fig8}, the MPANF outperforms the baseline methods across all four evaluation metrics. These empirical results align with our theoretical analysis and confirm the MPANF's consistent improvement over the naive baseline. This superior performance, even in challenging financial scenarios characterized by heteroscedasticity and high volatility, highlights the method's broad and reliable applicability.

\subsubsection{Retrospective Validation}

We further examined whether the condition expressed by Eq. \eqref{eq:15}, which specifies when the MPANF is expected to outperform the naive baseline in terms of MSE or RMSE, was satisfied across the eight target series. The purpose of this validation is to determine whether the theoretical conditions for superior performance are consistent with the actual results observed in the out-of-sample forecasts.

Table \ref{table:6} summarizes the validation results for each target series. Among the eight series where the MPANF achieved lower out-of-sample RMSE than the naive baseline, the retrospective condition held in all cases. Moreover, the in-sample coefficient $\hat{\alpha}^{*}_{\text{in}}$ is smaller than its out-of-sample counterpart $\hat{\alpha}^{*}_{\text{out}}$ in six of the eight series, indicating the robustness of the proposed method even when the movement-prediction accuracy decreased at the out of sample period.

\begin{table}[!t]
\centering
\caption{Retrospective validation for eight target series.}
\renewcommand{\arraystretch}{1.15}
\setlength{\tabcolsep}{5.5pt}
\begin{tabular}{lccccc}
\toprule
\textbf{Time series} & 
\(\hat{\alpha}^{*}_{\text{out}}\) & 
\(\bar{\epsilon}_{\text{out}}\) & 
\(\hat{\alpha}^{*}_{\text{in}}\) & 
\(\bar{\epsilon}_{\text{in}}\) & 
\textbf{Eq. \eqref{eq:15} Satisfied?} 
\\ 
\midrule
AAPL   & 0.0288 & 1.8963 & 0.1072 & 0.3842 6  & \checkmark \\
BA    & 0.0880 & 3.9675 & 0.1136 & 2.5867  & \checkmark \\
GOOGL  & 0.0240 & 1.5670 & 0.1008 & 0.4624   & \checkmark \\
GS & 0.1248 & 3.9736 & 0.1008 & 1.9030   & \checkmark \\
HAL & 0.1248 & 0.5426 & 0.1360 & 0.5252   & \checkmark \\
OIL    & 0.0880 & 1.2960 & 0.1152 & 0.9067   & \checkmark \\
SLB     & 0.1232 & 0.6905 & 0.1072 & 0.6596   & \checkmark \\
TSLA    & 0.0288 & 5.5773 & 0.1072 & 0.3534   & \checkmark \\
\bottomrule
\end{tabular}
\label{table:6}
\end{table}

\section{Discussion}
This section discusses the implications of the findings and the potential limitations of the study. We begin with the mechanism, interpretability, and potential value of the proposed method and then detail the conditions that constrain its performance.

Many real-world decision-making processes require combining heterogeneous forecasts such as binary labels and continuous values. Under this condition, MPANF provides a practical combination approach by translating movement predictions into point forecasts, enabling more informed decisions when directional signals are available and reliable. Its coefficient, which is a scaling factor, is derived in closed form under an unbiased movement prediction assumption. This analytical estimate can be regarded as a stable, regularized weight, which avoids noisy numerical tuning and makes the combination step lightweight, interpretable, and easy to deploy. Proposition 3 further states that MPANF improves on the naive baseline only if the out-of-sample product $\hat\alpha_{\text{out}}^{*} \cdot \bar{\epsilon}_{\text{out}}$ is at least half of the in-sample product $\hat\alpha_{\text{in}}^{*} \cdot \bar{\epsilon}_{\text{in}}$. In short, MPANF offers a clear mechanism for turning directional information into statistical gains, and it indicates a pathway via stronger movement prediction for potential economic relevance.

However, the effectiveness of MPANF is bottlenecked by the strength, bias, and stability of the directional signal. When directional signals are weak or unstable, MPANF may still underperform the naive baseline. In particular, if the assumption of an unbiased movement prediction does not hold, a substantial gap can arise between the analytical estimate and the empirical optimum, further diminishing the gains over the naive baseline. Absence of improvement should be diagnosed with Eq. \eqref{eq:15} to distinguish method limitations from genuine unpredictability and to guide remedies, such as developing more advanced movement prediction models. Finally, modest statistical gains may not translate into economic value after taking into account the transaction costs, slippage, and market frictions. Therefore, practical impact ultimately depends on directional signals that are both accurate and robust.

\section{Conclusions}
This study introduces the MPANF, a forecast combination method that adjusts the naive forecast by incorporating the movement prediction of the target variable. The proposed method demonstrates that systematic improvement over the naive baseline becomes theoretically and empirically possible when movement predictions are accurate and reliable. Thus, MPANF reframes the long-standing belief in the naive baseline’s supremacy: its apparent "unbeatability" in financial time series forecasting is not due to inherent optimality, but rather a consequence of the scarcity of reliable directional information. The dominance of the naive baseline, therefore, is a conditional property, not a universal one.

This study offers several directions for future research. First, since the current experiments employed movement predictions with modest accuracies, developing more advanced movement prediction models and evaluating the MPANF at higher accuracy levels would further demonstrate its effectiveness and enhance its practical value in real trading scenarios. Second, applying the MPANF to other types of time series beyond financial data, particularly those where the naive baseline is difficult to surpass but movement prediction is feasible, would broaden the applicability of the proposed method. Third, extending the MPANF to handle multivariate time series would enable testing its effectiveness in more complex, high-dimensional forecasting settings. Finally, while this study used a fixed adjustment factor for simplicity, developing adaptive adjustment schemes could further improve the method's performance.

\section*{Declarations}

\begin{itemize}
\item \textbf{Funding}\\
This study did not receive any specific grants from funding agencies in the public, commercial, or not-for-profit sectors.

\item \textbf{Conflict of interest/Competing interests}\\ 
The author reports that there are no competing interests to declare.

\item \textbf{Data availability}\\
The data are available at the following URL: \url{https://github.com/Zhang-Cheng-76200/MPANF}

\item \textbf{Code availability}\\
The code has been made available at the following URL: \url{https://github.com/Zhang-Cheng-76200/MPANF}

\item \textbf{Author contributions}\\
The author confirms being the sole contributor of this work.
\end{itemize}

\appendix
\section*{Appendix A. Algebraic Derivation of $\Delta MSE_{\text{in}}$} \renewcommand{\theequation}{A\arabic{equation}} \setcounter{equation}{0}

According to Eq. \eqref{eq:2}, the in-sample MSE of the naive forecast, denoted by $MSE_{\text{in}}^{\text{Naive}}$, can be expressed as:

\begin{align}
& MSE_{\text{in}}^{\text{Naive}} 
= \frac{1}{M} \left( \sum_{t \in \text{in}} \left( y_t - y_{t-1} \right)^2 \right) && \nonumber\\
&= \frac{1}{M} \left( \sum_{t \in \text{in}} \left( (y_{t-1} + d_t \cdot |\epsilon_t|) - y_{t-1} \right)^2 \right) && \nonumber\\
&= \frac{1}{M} \left( \sum_{t \in \text{correct}} |\epsilon_t|^2 + \sum_{t \in \text{incorrect}} |\epsilon_t|^2 \right)
\label{eq:A1}
\end{align}

By using Eq. \eqref{eq:2} and Eq. \eqref{eq:4} to replace $y_t$ and $y_{t-1}$ in Eq. \eqref{eq:A1}, respectively, the in-sample MSE of the MPANF, denoted by $MSE_{\text{in}}^{\text{MPANF}}$, can be expressed as follows: 

\begin{align}
& MSE_{\text{in}}^{\text{MPANF}} 
= \frac{1}{M} \left( \sum_{t \in \text{in}} \left( y_t - \hat{y}_t \right)^2 \right) && \nonumber\\
&= \frac{1}{M} \left( \sum_{t \in \text{correct}} \left( y_t - \hat{y}_t \right)^2 + \sum_{t \in \text{incorrect}} \left( y_t - \hat{y}_t \right)^2 \right) && \nonumber\\
&= \frac{1}{M} \left( \sum_{t \in \text{correct}} \left[ (y_{t-1} + d_t \cdot |\epsilon_t|) - (y_{t-1} + \hat{d}_t \cdot \alpha_{\text{in}} \cdot \bar{\epsilon}_{\text{in}}) \right]^2  \right. && \nonumber\\
&\quad \left. + \sum_{t \in \text{incorrect}} \left[ (y_{t-1} + d_t \cdot |\epsilon_t|) - (y_{t-1} + \hat{d}_t \cdot \alpha_{\text{in}} \cdot \bar{\epsilon}_{\text{in}}) \right]^2 \right) && \nonumber\\
&= \frac{1}{M} \left( \sum_{t \in \text{correct}} \left( |\epsilon_t|^2 - 2 \cdot \alpha_{\text{in}} \cdot |\epsilon_t| \cdot \bar{\epsilon}_{\text{in}} + \alpha_{\text{in}}^2 \cdot \bar{\epsilon}_{\text{in}}^2 \right) \right. && \nonumber\\
&\quad \left. + \sum_{t \in \text{incorrect}} \left( |\epsilon_t|^2 + 2 \cdot \alpha_{\text{in}} \cdot |\epsilon_t| \cdot \bar{\epsilon}_{\text{in}} + \alpha_{\text{in}}^2 \cdot \bar{\epsilon}_{\text{in}}^2 \right) \right) 
\label{eq:A2}
\end{align}

The in-sample MSE difference between the naive forecast and the MPANF, denoted by $\Delta MSE_{\text{in}}$, can subsequently be derived via Eqs. \eqref{eq:A1} and \eqref{eq:A2}: 

\begin{align}
& \Delta MSE_{\text{in}} 
= MSE_{\text{in}}^{\text{Naive}} - MSE_{\text{in}}^{\text{MPANF}} && \nonumber\\
&= \frac{1}{M} \left( \sum_{t \in \text{correct}} |\epsilon_t|^2 + \sum_{t \in \text{incorrect}} |\epsilon_t|^2 \right)  && \nonumber\\
&\quad - \frac{1}{M} \left( \sum_{t \in \text{correct}} \left( |\epsilon_t|^2 - 2 \cdot\alpha_{\text{in}} \cdot|\epsilon_t| \cdot\bar{\epsilon}_{\text{in}} + \alpha_{\text{in}}^2\cdot \bar{\epsilon}_{\text{in}}^2 \right) \right. && \nonumber\\
&\quad \left. + \sum_{t \in \text{incorrect}} \left( |\epsilon_t|^2 + 2 \cdot\alpha_{\text{in}}\cdot |\epsilon_t| \cdot \bar{\epsilon}_{\text{in}} + \alpha_{\text{in}}^2 \cdot \bar{\epsilon}_{\text{in}}^2 \right) \right) && \nonumber\\
&= \frac{1}{M} \left( \sum_{t \in \text{correct}} \left( 2 \cdot\alpha_{\text{in}} \cdot|\epsilon_t| \cdot\bar{\epsilon}_{\text{in}} - \alpha_{\text{in}}^2 \cdot \bar{\epsilon}_{\text{in}}^2 \right) - \sum_{t \in \text{incorrect}} \left( 2 \cdot\alpha_{\text{in}} \cdot|\epsilon_t| \cdot \bar{\epsilon}_{\text{in}} 
+ \alpha_{\text{in}}^2 \cdot \bar{\epsilon}_{\text{in}}^2 \right) \right) 
\label{eq:A3}
\end{align}

Assuming that the in-sample movement prediction is unbiased and contains sufficient numbers of both correct and incorrect predictions, the following approximations hold on the basis of the law of large numbers (LLN):

\begin{equation} 
\label{eq:A4} 
\sum_{t \in \text{correct}} |\epsilon_t| \approx m_{\text{in}}^{\text{correct}} \cdot \bar{\epsilon}_{\text{in}}, \quad
\sum_{t \in \text{incorrect}} |\epsilon_t| \approx m_{\text{in}}^{\text{incorrect}} \cdot \bar{\epsilon}_{\text{in}}
\end{equation}

Therefore, Eq. \eqref{eq:A3} can be simplified as follows:: 

\begin{align}
& \Delta MSE_{\text{in}} 
= MSE_{\text{in}}^{\text{Naive}} - MSE_{\text{in}}^{\text{MPANF}} && \nonumber\\
&\approx \frac{1}{M} \left( 2\cdot \alpha_{\text{in}} \cdot \bar{\epsilon}_{\text{in}} \cdot (m_{\text{in}}^{\text{correct}} \cdot \bar{\epsilon}_{\text{in}}) - m_{\text{in}}^{\text{correct}} \cdot \alpha_{\text{in}}^2 \cdot \bar{\epsilon}_{\text{in}}^2  \right. && \nonumber\\
&\quad \left.- 2\cdot \alpha_{\text{in}} \cdot \bar{\epsilon}_{\text{in}} \cdot (m_{\text{in}}^{\text{incorrect}} \cdot \bar{\epsilon}_{\text{in}}) - m_{\text{in}}^{\text{incorrect}} \cdot \alpha_{\text{in}}^2 \cdot \bar{\epsilon}_{\text{in}}^2 \right) && \nonumber\\
&= \left( 4\cdot \alpha_{\text{in}} \cdot ACC_{\text{in}} - \alpha_{\text{in}}^2 - 2\cdot \alpha_{\text{in}} \right) \cdot \bar{\epsilon}_{\text{in}}^2 
\label{eq:A5}
\end{align}

\section*{Appendix B. Algebraic Derivation of $\Delta MSE_{\text{out}}$} \renewcommand{\theequation}{B\arabic{equation}} \setcounter{equation}{0}

When $\hat\alpha_{\text{in}}^{*}$ and $\bar{\epsilon}_{\text{in}}$ are adopted by the MPANF on the out-of-sample set, the out-of-sample MSE difference between the naive forecast and the MPANF, denoted by $\Delta MSE_{\text{out}}$, can be expressed as follows:

\begin{align}
& \Delta MSE_{\text{out}} 
= MSE_{\text{out}}^{\text{Naive}} - MSE_{\text{out}}^{\text{MPANF}} && \nonumber\\
&= \frac{1}{N} \left( \sum_{t \in \text{correct}} |\epsilon_t|^2 + \sum_{t \in \text{incorrect}} |\epsilon_t|^2 \right) && \nonumber\\
&\quad - \frac{1}{N} \left( \sum_{t \in \text{correct}} \left( |\epsilon_t|^2 - 2 \cdot\hat\alpha_{\text{in}}^{*} \cdot|\epsilon_t| \cdot\bar{\epsilon}_{\text{in}} + \left(\hat\alpha_{\text{in}}^{*}\right)^2\cdot \bar{\epsilon}_{\text{in}}^2 \right) \right. && \nonumber\\
&\quad \left. + \sum_{t \in \text{incorrect}} \left( |\epsilon_t|^2 + 2 \cdot\hat\alpha_{\text{in}}^{*}\cdot |\epsilon_t| \cdot\bar{\epsilon}_{\text{in}} + \left(\hat\alpha_{\text{in}}^{*}\right)^2 \cdot\bar{\epsilon}_{\text{in}}^2 \right) \right) && \nonumber\\
&= \frac{1}{N} \left( \sum_{t \in \text{correct}} \left( 2 \cdot\hat\alpha_{\text{in}}^{*} \cdot|\epsilon_t|\cdot \bar{\epsilon}_{\text{in}} - \left(\hat\alpha_{\text{in}}^{*}\right)^2 \cdot\bar{\epsilon}_{\text{in}}^2 \right) - \sum_{t \in \text{incorrect}} \left( 2\cdot \hat\alpha_{\text{in}}^{*} \cdot|\epsilon_t| \cdot\bar{\epsilon}_{\text{in}} + \left(\hat\alpha_{\text{in}}^{*}\right)^2 \cdot\bar{\epsilon}_{\text{in}}^2 \right) \right)
\label{eq:B1}
\end{align}

Similarly, assuming that the out-of-sample movement prediction is unbiased and contains sufficient numbers of both correct and incorrect predictions, the following approximations hold on the basis of the LLN:

\begin{equation} 
\label{eq:B2} 
\sum_{t \in \text{correct}} |\epsilon_t| \approx n_{\text{out}}^{\text{correct}} \cdot \bar{\epsilon}_{\text{out}}, \quad
\sum_{t \in \text{incorrect}} |\epsilon_t| \approx n_{\text{out}}^{\text{incorrect}} \cdot \bar{\epsilon}_{\text{out}}
\end{equation}

Therefore, Eq. \eqref{eq:B1} can further be expressed as: 

\begin{align}
& \Delta MSE_{\text{out}}
= MSE_{\text{out}}^{\text{Naive}} 
 - MSE_{\text{out}}^{\text{MPANF}} && \nonumber\\
& \approx \frac{1}{N} \left(2\cdot \hat\alpha_{\text{in}}^{*} \cdot \bar{\epsilon}_{\text{in}} \cdot (n_{\text{out}}^{\text{correct}} \cdot \bar{\epsilon}_{\text{out}}) \right. - n_{\text{out}}^{\text{correct}} \cdot (\hat\alpha_{\text{in}}^{*})^2 \cdot \bar{\epsilon}_{\text{in}}^2 && \nonumber\\
& \quad \left. - 2\cdot \hat\alpha_{\text{in}}^{*} \cdot \bar{\epsilon}_{\text{in}} \cdot (n_{\text{out}}^{\text{incorrect}} \cdot \bar{\epsilon}_{\text{out}}) - n_{\text{out}}^{\text{incorrect}} \cdot (\hat\alpha_{\text{in}}^{*})^2 \cdot \bar{\epsilon}_{\text{in}}^2 \right) && \nonumber\\
&= 2\cdot \hat\alpha_{\text{in}}^{*} \cdot \bar{\epsilon}_{\text{in}} \cdot ACC_{\text{out}} \cdot \bar{\epsilon}_{\text{out}} - 2\cdot \hat\alpha_{\text{in}}^{*} \cdot \bar{\epsilon}_{\text{in}}  \cdot (1-ACC_{\text{out}}) \cdot \bar{\epsilon}_{\text{out}} - (\hat\alpha_{\text{in}}^{*})^2 \cdot \bar{\epsilon}_{\text{in}}^2 && \nonumber\\
&\approx 2\cdot \hat\alpha_{\text{in}}^{*} \cdot \bar{\epsilon}_{\text{in}} \cdot \hat\alpha_{\text{out}}^{*} \cdot \bar{\epsilon}_{\text{out}} - (\hat\alpha_{\text{in}}^{*})^2 \cdot \bar{\epsilon}_{\text{in}}^2
\label{eq:B3}
\end{align}

\section*{Appendix C. Movement Prediction Using Co-moving Variables} \renewcommand{\theequation}{C\arabic{equation}} \setcounter{equation}{0}
A simple but effective method for generating movement prediction involves leveraging the co-movement between the target variable and an exogenous variable. The selection of the exogenous variable is typically guided by domain knowledge, and the movement prediction of the target variable, denoted by $\hat{d}_{t}^{\text{target}}$, can be simply obtained as:
\begin{equation}
\label{eq:C1}
\hat{d}_{t}^{\text{target}} = \hat{d}_{t}^{\text{exo}},
\end{equation}
where $\hat{d}_{t}^{\text{exo}}$ represents the movement prediction of the exogenous variable at time step $t$.
In particular, if the value of the exogenous variable is recorded sufficiently ahead of the target variable for the same time step, such as in financial markets operating in different time zones, the actual movement of the exogenous variable can directly serve as the movement prediction for the target variable:
\begin{equation}
\label{eq:C2}
\hat{d}_{t}^{\text{target}} = d_{t}^{\text{exo}}=
\begin{cases} 
      1, & \text{if } y_{t}^{\text{exo}} > y_{t-1}^{\text{exo}}; \\ 
      -1, & \text{if } y_{t}^{\text{exo}} \leq y_{t-1}^{\text{exo}};
\end{cases},
\end{equation}
where $d_{t}^{\text{exo}}$ is the actual movement of the target variable at time step $t$. $y_{t}^{\text{exo}}$ and $y_{t-1}^{\text{exo}}$ denotes the actual values of the exogenous variable at time steps $t$ and $t-1$, respectively. 
\bibliographystyle{elsarticle-harv} 
\bibliography{references}



\end{document}